\documentclass[11pt,superscriptaddress]{revtex4-1}
\usepackage{adjustbox}
\usepackage{subfig}
\usepackage{wrapfig}
\usepackage[usenames,dvipsnames]{color}
\usepackage{amssymb}
\usepackage{amsmath}
\usepackage{enumerate}
\usepackage{graphicx}
\usepackage{bm}
\usepackage{hyperref}
\usepackage{makeidx}
\usepackage{lineno}
\usepackage{subfig} 
\usepackage{booktabs}
\usepackage{subfig}
\usepackage{wrapfig}
\usepackage{adjustbox}

\begin{document}

\title{Experimental and numerical study of the effect of surface patterning on the frictional properties of polymer surfaces}

\author{Simone Balestra}
\affiliation{\footnotesize Department of Physics and Nanostructured Interfaces and Surfaces Centre, University of Torino, Via Pietro Giuria 1, 10125, Torino, Italy}
\email{simone.balestra@itt.com \looseness=-1}	

\author{Gianluca Costagliola} 
\affiliation{\footnotesize Civil Engineering Institute, Materials Science and Engineering Institute, Ecole Polytechnique Federale de Lausanne - EPFL, 1015 Lausanne, Switzerland \looseness=-1}
\email{gianluca.costagliola@epfl.ch}

\author{Amedeo Pegoraro}
\affiliation{\footnotesize Department of Physics and Nanostructured Interfaces and Surfaces Centre, University of Torino, Via Pietro Giuria 1, 10125, Torino, Italy\looseness=-1} 

\author{Federico Picollo} 
\affiliation{\footnotesize Department of Physics and Nanostructured Interfaces and Surfaces Centre, University of Torino, Via Pietro Giuria 1, 10125, Torino, Italy\looseness=-1} 

\author{Jean-Fran\c{c}ois Molinari}
\affiliation{\footnotesize Civil Engineering Institute, Materials Science and Engineering Institute, Ecole Polytechnique Federale de Lausanne - EPFL, 1015 Lausanne, Switzerland \looseness=-1}

\author{Nicola M. Pugno} 
\affiliation{\footnotesize Laboratory of Bio-inspired, Bionic, Nano, Meta Materials \& Mechanics, Department of Civil, Environmental and Mechanical Engineering, University of Trento, Via Mesiano, 77, 38123 Trento, Italy \looseness=-1}
\affiliation{\footnotesize School of Engineering and Materials Science, Queen Mary University of London, Mile End Road, London E1 4NS, UK\looseness=-1}

\author{Ettore Vittone} 
\affiliation{\footnotesize Department of Physics and Nanostructured Interfaces and Surfaces Centre, University of Torino, Via Pietro Giuria 1, 10125, Torino, Italy\looseness=-1} 

\author{Federico Bosia }
\affiliation{\footnotesize DISAT, Politecnico di Torino, Corso Duca degli Abruzzi 24, 10129 Torino, Italy\looseness=-1}

\author{Agusti Sin }
\affiliation{\footnotesize ITT motion technologies, Via Molini 19, 12032 Barge, Italy\looseness=-1}

\begin{abstract}
We describe benchmark experiments to evaluate the frictional properties of laser patterned low-density polyethylene as a function of sliding velocity, normal force and humidity. The pattern is a square lattice of square cavities with sub-mm spacing. We find that dynamic friction decreases compared to non-patterned surfaces, since stress concentrations lead to anticipated detachment, and that stick-slip behavior is also affected. Friction increases with humidity, and the onset of stick-slip events occurs in the high humidity regime. Experimental results are compared with numerical simulations of a simplified 2-D spring-block model. A good qualitative agreement can be obtained by introducing a deviation from the linear behavior of the Amontons-Coulomb law with the load, due to a saturation in the effective contact area with pressure. This also leads also to the improvement of the quantitative results of the spring-block model by reducing the discrepancy with the experimental results, indicating the robustness of the adopted simplified approach, which could be adopted to design patterned surfaces with controlled friction properties.
\end{abstract}

\maketitle  

\section{Introduction}

Surface morphology plays an essential role in determining friction properties in many engineering applications. For instance, in the automotive field (e.g. tires, brake pads, bearings and related technology), friction is responsible for energy losses and material consumption \cite{Erdemir1}\cite{Erdemir2}, so that understanding the effects of the surface morphology on the emergent behavior is a fundamental requirement to improve the efficiency and boost environmental sustainability \cite{Priest}. 

Recent studies suggest that this can be achieved by means of an appropriate surface design of automotive components \cite{Ryk} \cite{Ferreira}. The underlying idea is that macroscopic friction properties can be modified by means of surface microstructures , e.g. artificial patterns allowing a high tunability of tribological performance \cite{Rosenkranz}, although it is difficult to practically exploit these results to find optimized solutions, due to the variety and the complexity of materials and tribological systems \cite{Gachot}\cite{Gropper}. Useful insights to manipulate friction have been obtained with hexagonal structures \cite{Varenberg}\cite{Li}, arrays of cavities  \cite{He}\cite{Berardo} and grooves \cite{Baum}.
Laser texturing has been shown to be an efficient technique to tune frictional properties, as demonstrated in various studies. Etsion et al have adopted this technique for mechanical seals \cite{EtsionHalperin}, Greiner et al. to modify static friction \cite{Greiner}, Hsu et al discussed its relevance in fatigue phenomena \cite{Hsu} and Gnilitskyi et al. have applied it to metal surfaces \cite{Gnilitskyi}.
A review of the state of the art is provided in \cite{Etsion}\cite{Fiaschi}.
In particular, Maegawa et al. \cite{Maegawa} studied the effect on kinetic friction of the number of surface grooves present on contact surfaces and found that friction decreased as the number of grooves increased. Capozza et  al. \cite{Capozza} indicated that macroscopic surface grooves are theoretically effective in reducing static friction, due to a non-uniform distribution of surface stress induced by patterning \cite{Costagliola2}. Tunability can be achieved by means of a hierarchical organization \cite{Costagliola1} or combining various types of microstructures \cite{Costagliola0}.

In addition to the problem of friction-related energy dissipation, stick-slip phenomena in sliding friction can generate mechanical vibrations leading to noise pollution \cite{diLiberto}. Surface modifications of the topology can reduce or eliminate this effect by varying the static and dynamic friction coefficients without changing the chemical composition of the surface. Surface grooves can reduce this type of noise, as demonstrated by Wang \cite{Wang}.

However, friction surfaces operate in different environmental conditions, so that the effect of patterning also needs to be evaluated in the presence of varying temperature and humidity. Bhushan et al. \cite{Bhushanart} have shown that there are different regimes as a function of the amount of water present between the surfaces. Hence, if the surface geometry changes, e.g. in the presence of grooves or cavities, regimes can vary with the same amount of water. Meniscus forces strongly depend on the distribution of asperities and the amount of water \cite{Bhushanlibro}.

In this paper, we study the effect of the modification of surface morphology at the micrometer scale on kinetic friction, using laser patterning to create periodic arrays of cavities on a polyethylene sample. We focus the frictional behavior induced by this patterning as a function of several external parameters, e.g. sliding velocity, normal force and relative humidity, from the dry case to the high humidity regime. Friction force is measured by a tribometer and surface morphology is assessed using a profilometer. Finally, experimental data are qualitatively compared with the numerical predictions of a simplified 2-D spring-block model. Overall, we propose a method to determine the effective frictional properties of arbitrarily structured patterned polymer surfaces.


\section{Materials and methods}\label{sec2}

\subsection{Specimens}

The chosen material for the samples is low-density polyethylene due to its well-known mechanical properties and good machinability, as well as its limited wear effects on its tribological counterpart made of gray cast iron.
Two types of polyethylene samples have been tested: the first is nominally flat and the second is patterned. The geometry of the samples is circular, both have a diameter of 1 cm and a thickness of 0.7 cm. The Young’s Modulus of the material is (0.38$\pm$0.08) GPa.

\subsection{Laser patterning}

\begin{figure}
\centering
\includegraphics[width=0.55\linewidth]{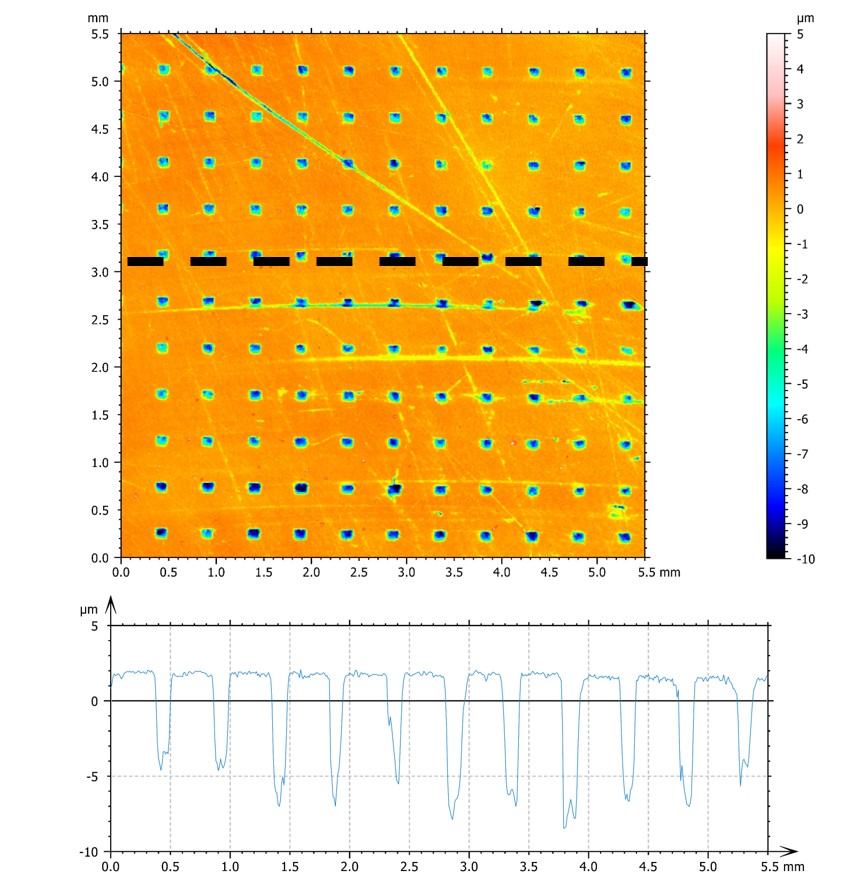}
\caption{top - optical images of the patterned surface. bottom - depth profile along the black line shown in the optical image (depth profile extracted at height y of 3 mm). To the right of the color scale is the histogram of the height distribution.}
\label{fig:ProfilometroPatternBeforeDash}
\end{figure}

Surface micro-patterning was performed with a ns pulsed Nd:YAG laser (EzLaze3 by New  Wave); the laser wavelength is in the infrared range ($\lambda=1064$ nm). The pattern is a lattice of square cavities whose side is (130$\pm$8) $\mu$m long. The distance between cavities is (370$\pm$13) $\mu$m in x and y directions. Using two bursts for each cavity, the obtained depth is (5.2$\pm$1.3) $\mu$m. The pulse duration is 4 ns, the spot size is 23 $\mu$m and the power density is 3.65 GW/cm$^{2}$. Figure \ref{fig:ProfilometroPatternBeforeDash} shows the altitude map of the patterned surface and the depth profile obtained by an optical profilometer. Altitude maps of other samples are reported in the Supplementary Material \ref{sec_suppl}.
Since the numerical model that we use in this work introduces a square discretization, square cavities can be simulated without shape approximations. Length and spacings of the patterning have been selected as a compromise between minimizing the number of laser pulses and the requirement to texture the whole surface with a large number of cavities with a sufficient depth and maximizing the number of cavities with a sufficient depth over the whole surface.

\subsection{Tribometer}

\begin{figure}
    \centering
    \begin{tabular}{cc}
    \adjustbox{valign=b}{\subfloat[\label{subfig-1:sh}]{%
          \includegraphics[width=.4\linewidth,height=7.1cm]{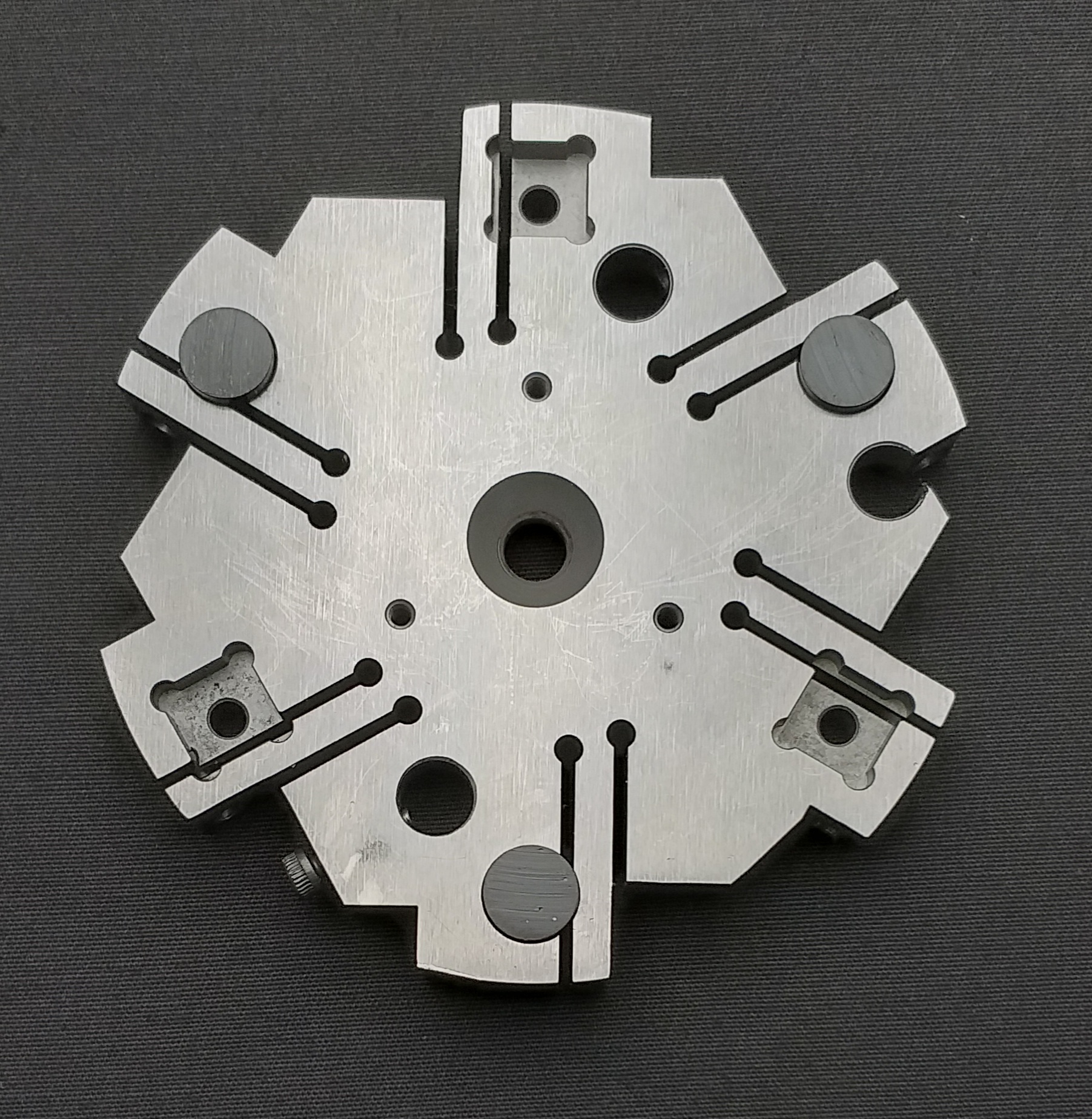}}}
    &      
    \adjustbox{valign=b}{\begin{tabular}{@{}c@{}}
    \subfloat[\label{subfig-2:fl}]{%
          \includegraphics[width=.3\linewidth]{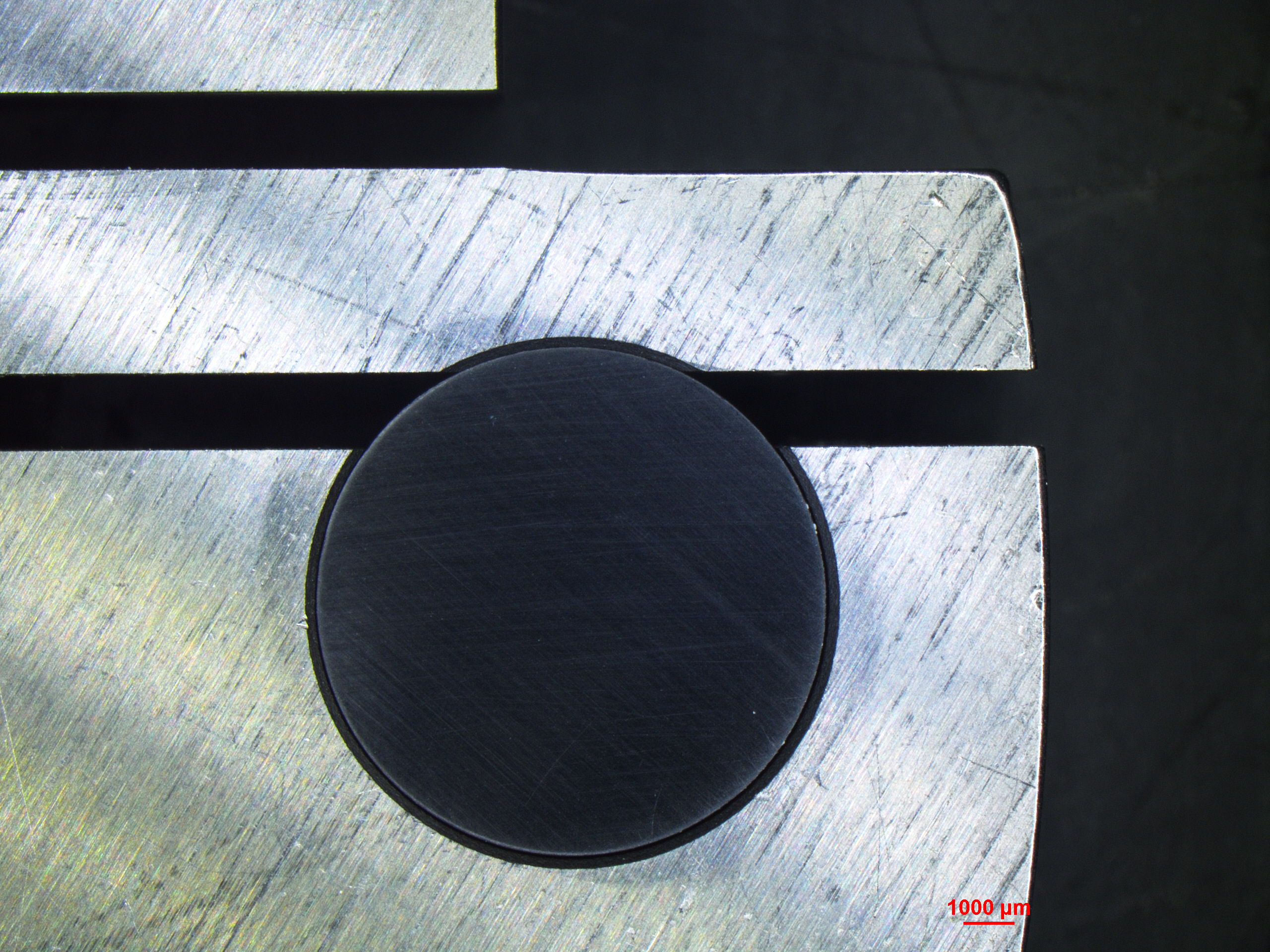}} \\
    \subfloat[\label{subfig-3:pt}]{%
          \includegraphics[width=.3\linewidth]{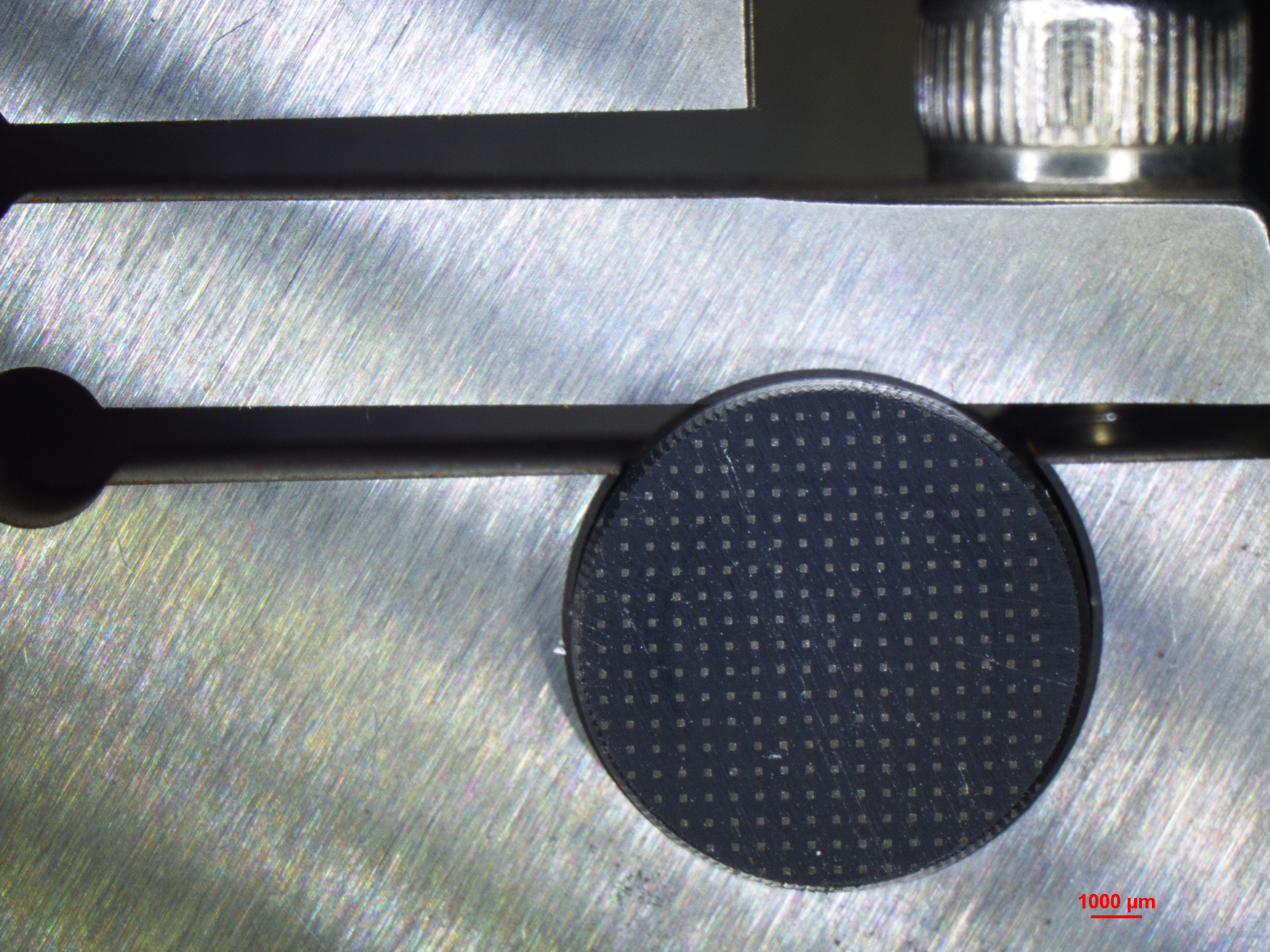}}
    \end{tabular}}
    \end{tabular}
    \caption{Tribometer sample holder (a), flat (b) and patterned (c) polyethylene sample fixed on the tribometer sample holder. The effective radius is $r_{eff}=38.1$ mm. }\label{fig:SampleHolder}
  \end{figure}

Friction coefficients are measured using a Bruker UMT-TriboLab tribometer \cite{TriboLab}. The sample holder can hold three identical samples, in order to better distribute the pressure (figure \ref{fig:SampleHolder}). In the case of patterned samples, the sliding direction is parallel to the sides of the cavities. The measurement was performed using a cast iron disk, which does not change its roughness during the test because the elastic modulus of iron is larger than polyethylene. 

From the measurement of the time evolution of the torque $T_z(t)$, given the normal force $F_n(t)$ acting on the sample and the effective radius $r_{eff}$ (the effective  radius  is the  distance  between the center  of the  sample holder  and the center  of  the sample  position, considering it point-like), the coefficient of friction ($\mu$) is calculated through the following formula:

\begin{equation}
\mu(t)=\frac{T_z(t)}{F_n(t) \cdot{r_{eff}}}
\label{eq.TribCoF1}
\end{equation}

This definition is consistent with the conventional one, i.e. the ratio between tangential and normal force. Since the tribometer setup is designed to measure a torque, the tangential force can be obtained by dividing the torque by the distance of the samples from the center of the tribometer.
The tribometer can work in a climatic chamber that allows operations with a controlled relative humidity (RH) ranging from 10 \% to 90 \% . The torque sensor works using a Wheatstone bridge fixed on a shock absorber that isolates it from external vibrations. Friction coefficients were measured for flat and patterned samples as a function of sliding velocity, relative humidity and normal force.

\subsection{Optical profilometer}

Surface morphology and roughness was characterized by a Nanovea PS50 profilometer \cite{Profilometer} based on the Chromatic Confocal technique, which provides measurements with nanometric and micrometric depth and lateral resolution, respectivery, and a scan area of 5x5 cm$^{2}$.

\section{Theory and calculation}\label{sec3}

To support the interpretation of experimental results by means of a simplified numerical approach, we adopt the spring-block model \cite{burr}, which has been already used to investigate frictional phenomena \cite{urbakh}\cite{trom}. In \cite{Costagliola2}, the model was implemented in a 2-D formulation to model the horizontal contact plane and study the effects on friction due to the surface pattern geometry. This formulation is particularly useful in investigating static friction and the behavior at the onset of sliding, but is less effective in describing dynamic friction. In particular, results in \cite{Costagliola2} for the dynamic friction coefficient of surfaces with square cavity arrays do not match the observed experimental results in this work for the dry case (e.g. section \ref{sec_exp_res}). In order to extract quantitative predictions from the spring-block model, a validation of the model is required through a comparison of the experimental results and the simulated outcomes.

For this reason, we consider instead a 2-D discretization of the substrate in the vertical plane, similarly to the approach adopted in \cite{trom}, in order to take into account the vertical stress distribution. Since the presence of surface structures induces stress concentrations, we expect that this may play a role in determining the dynamic friction behavior. Although this approximation neglects the 2-D pattern geometry on the horizontal contact plane, we first adopt this simplified numerical model to avoid a computationally expensive 3-D formulation. Thus, the cavities are approximated along their vertical profile shown in figure \ref{fig:ProfilometroPatternBeforeDash}.

\begin{figure}[h!]
\hspace{5cm}
\includegraphics[scale=0.5]{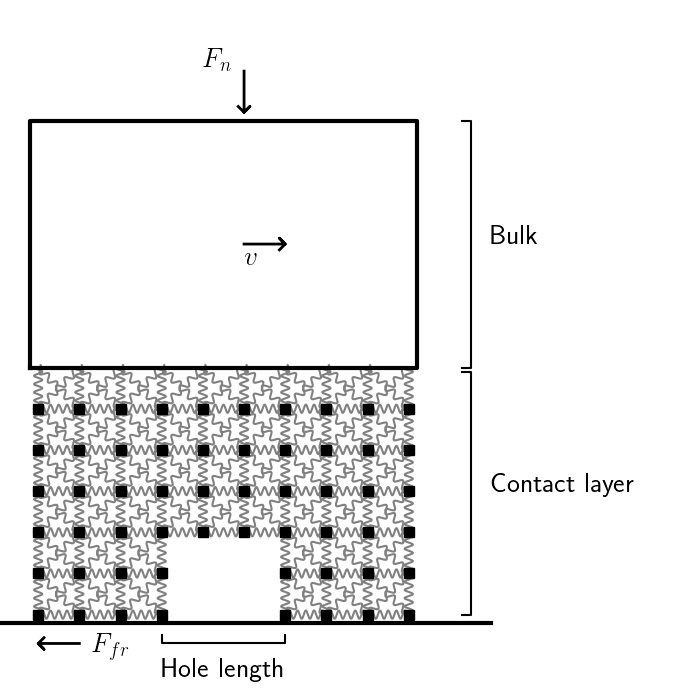}
\caption{Schematic of the 2-D spring-block model used to approximate the experimental apparatus. Note that length and depth of the cavity in this scheme do not correspond to the adopted lengths.}
\label{fig:Model}
\end{figure}

The system is modelled as follows: we discretize the lower layer of the sample by means of a 2-D spring-block square mesh, in which each block is attached to its eight neighbors by means of linear springs, while the remaining bulk portion is considered as a single rigid block, as shown in figure \ref{fig:Model}. We assume that in the pressure and velocity ranges considered experimentally, the effects due to this bulk approximation are negligible. A uniform pressure $P$ is applied along the vertical axis to the bulk, moving at constant velocity $v$ in the horizontal direction. The system slides over a rigid undeformable surface, reproducing the geometry of the experimental apparatus of a patterned polyethylene sample sliding over cast iron. 

We fix the dicretization length to $l= 10 \; \mu$m, which is small enough to take into account the cavity depth and large enough to introduce an effective friction law on an elementary surface unit represented by the block. The qualitative behavior is not affected by this value. Given the density of polyethylene ($\rho=0.9$ g/cm$^3$), the mass of each block is $m=\rho l^3$ where $l$ is the side of the square, while it is fixed to $m/2$ for blocks located on edges of the mesh and $m/4$ for those on the corners. The stiffness can be fixed by imposing the equivalence with an isotropic elastic material with Young's modulus $E$ (for polyethylene, $E=0.38$ GPa) with the procedure illustrated in \cite{absi}. Thus, the stiffness is $k=3/4 E l$ and $k/2$ for diagonal springs. In this mesh, the value of the Poisson's ratio is fixed to $\nu=1/3$, which is reasonably close to the realistic value.

The internal elastic force applied to a block $i$ by its neighbor $j$ is $\textbf{F}_{int}^{(i,j)} = k_{ij} (r_{ij}-l_{ij}) (\textbf{r}_j-\textbf{r}_i)/r_{ij} $, where $\textbf{r}_i$, $\textbf{r}_j$ are the position vectors, $r_{ij}$ is the modulus of their distance, $k_{ij}$ is the stiffness of the spring connecting them and $l_{ij}$ is the modulus of their rest distance. The springs between the upper row of blocks and the rigid bulk portion transmit the forces to all underlying blocks due to the normal pressure and the velocity imposed to the bulk portion. 

Each block is subjected to viscous damping to avoid artificial oscillations, $\textbf{F}_{d}^{(i)} = - 
\gamma m \dot{\textbf{r}_i}$, where $\gamma$ is the damping frequency. We fix $\gamma=10^{-2}\sqrt{k/m}$ in the underdamped regime. This choice does not affect qualitative results \cite{trom}.

Thus, the only forces applied to the blocks are the elastic forces due to the neighbors and the damping, except for those located on the bottom row, which are subjected to a friction force. This layer is constrained to horizontal motion along the substrate. In any case, we have verified that relaxing this condition does not lead to detachment with the chosen parameters.

A local static and a dynamic friction coefficient is assigned to each block $i$ in contact with the substrate, namely $\mu_s^{i}$ and $\mu_d^{i}$, respectively. These are randomly extracted at the beginning of the simulation from a Gaussian distribution with means $(\mu_s)_m$, $(\mu_d)_m$ and standard deviations $(\sigma_s)$, $(\sigma_d)$. This implies a coarse-grained description  with an effective friction law at a mesoscale that is larger than surface roughness. Statistical fluctuations of the roughness are taken into account by statistical distribution.

In the standard formulation, the local friction force is determined by the classical Amontons-Coulomb (AC) friction force, i.e. a force proportional to the total normal forces acting on $i$ due to its neighbors. In symbols, the vertical component of the total internal elastic force on $i$ is $F_n^{i} \equiv \sum_j F_{int}^{ij}|_y $. Thus, if a block is at rest, the static friction force is opposite to the horizontal forces acting on $i$, i.e. $F_{fr}= - \sum_j F_{int}^{ij}|_x$, up to the threshold
 $F_{fr}=\mu_s^{i} F_n^{i}$. If the block is moving, the dynamic friction force is $F_{fr}^{i}=\mu_d^{i}F_n^{i}$ in modulus and opposite to its velocity.

As discussed below, this simple AC friction force is inadequate to reproduce the experimental results and needs to be modified. Whatever friction force $\textbf{F}_{fr}^{(i)}$ is adopted, we can write the Newton's law for a block $i$: 
$m \ddot{\textbf{r}_i} = \sum_j \textbf{F}_{int}^{(ij)}+\textbf{F}_{fr}^{(i)} + \textbf{F}_{d}^{(i)}$.
By numerically integrating the whole system of equations with a fourth-order Runge-Kutta algorithm, the time evolution and all the physical quantities can be calculated. An elementary time step $dt=10^{-9}$ s is sufficient to avoid integration errors.

Simulations are actually performed in two steps: first we apply the normal force with velocity $v=0$, increasing the force from zero to the chosen value in a short but finite time. This avoids numerical instabilities due to the abrupt force application. We have found that increasing the force in an interval of $2 \cdot 10^{-6}$ s is sufficient for this purpose. Then, the force is kept constant for further $10^{-5}$ s, which are sufficient  to damp internal oscillations and dissipate the kinetic energy of the mesh. After the system is stabilized, the velocity is switched to the experimental value.

Pressure and velocity are the same as in the experiments, with $v=2$ mm/s and pressure between $P=0.65$ MPa corresponding to the experimental case of 50 N, and $P=2.6$ MPa for the case of 200 N. The number of blocks are fixed by the elementary length $l$ and the real experimental specimen length: the height of the contact layer is fixed to 20 blocks, corresponding to 200 $\mu$m , which is sufficiently large compared with the cavity depth. The length of the system is 996 blocks, in order to approximately match the experimental samples length of 1 cm. In the patterned case, the length of each cavity in the mesh is 130 $\mu$m and the depth is fixed to 20 $\mu$m. 

The local friction coefficients are tuned to obtain a macroscopic behavior comparable with the experimental one for the nominally flat samples and, in particular, to exactly match the dynamic friction coefficient. Then, with these same parameters, we introduce the cavities, to verify the validity of results for the patterned case. Thus, the set of four variables determining the local distribution has been adjusted once and for all in preliminary tests for the non-patterned case. These values depend on the selected friction law, which is discussed in the next section.

\subsection{Calculation of the effective contact area}

\begin{figure}[h!]
\begin{center}
\includegraphics[scale=0.4]{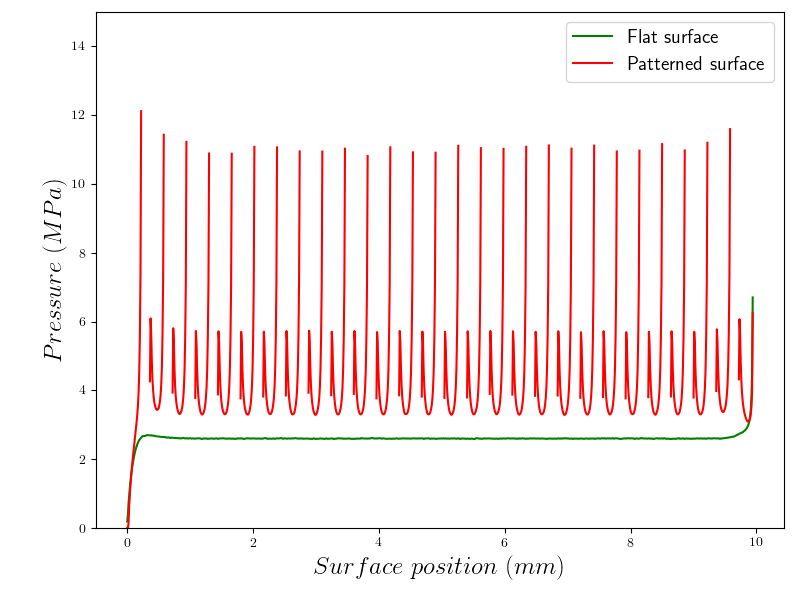}
\caption{Comparison of the local pressure distribution for patterned and flat case during the sliding with a total normal force applied of $200$ N.  }
\label{fig:Pressure}
\end{center}
\end{figure}

In theoretical and numerical models, the effective contact area is usually assumed proportional to the applied load, leading to the classical AC friction law \cite{BowdenTabor}. However, in the case of the patterning, there is a non-uniform vertical stress distribution on the contact surface characterized by concentrations located at each cavity edge, as shown in figure \ref{fig:Pressure}. Thus, in the flat case, stress concentrations are only located at the edges, while in the patterned case they occur all along the surface. 

For this reason, we expect that, in the presence of corrections to the linear behavior of the contact area with the pressure, these will affect much more the friction of the patterned surface than of the flat one, leading to an effect on the dynamic friction observed experimentally. To verify this scenario, we investigate the behavior of the effective contact area with the load for our polyethylene sample.

We consider the surface profile of the flat sample before the test reported in the Supplemental Material \ref{sec_suppl} (figure \ref{profilometroFlat}). In order to calculate the effective contact area with the load, we use the software TAMAAS \cite{tamaas}, a freely available high-performance code based on boundary and volume integral equations \cite{frerot}. We assume a model of linear elasticity for the sample, while the only external parameters are the Young's modulus $E$, the Poisson's ratio $\nu$ and the sample length. The software calculates the effective contact area as a function of the applied pressure $P$.

\begin{figure}[h!]
\begin{center}
\includegraphics[scale=0.4]{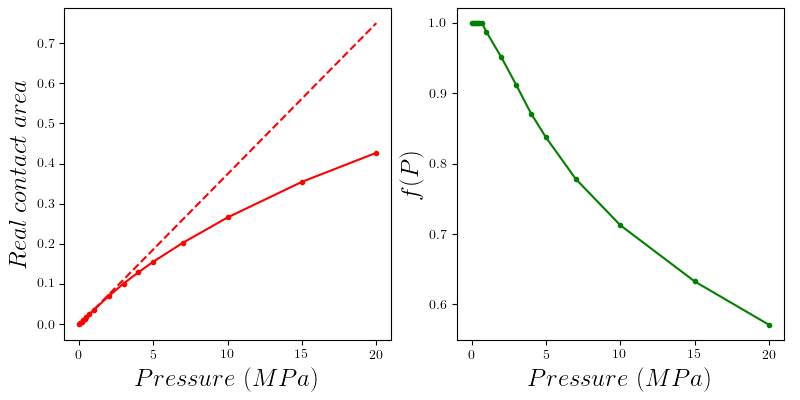}
\caption{Left: Effective contact area as a function of pressure obtained with TAMAAS \cite{tamaas} for the flat sample surface of figure \ref{profilometroFlat}. The dashed line represents a linear behavior starting from the chosen threshold 0.7 MPa. Right: correction factor $f(P)$ to modify the AC local friction law.}
\label{fig:AreaFrac}
\end{center}
\end{figure}

Results for the nominally flat sample are shown in the left panel of figure \ref{fig:AreaFrac}: a deviation from a linear behavior is observed for increasing pressure, in particular the curve saturates for pressures whose order of magnitude corresponds to that estimated with the model. These results are not affected by the scale of the sample, i.e. repeating the test for smaller portions of the surface. Given this result, we can assume that the effective contact area corresponding to our elementary block does not scale linearly with the load, but saturates for larger pressures. Thus, the standard friction force in the spring block model must be modified.

We therefore assume a linear AC friction law up to a threshold corresponding to the smallest experimental pressure: from this we assume a law proportional to the effective contact area.
In symbols, given the static or dynamic friction coefficient, the friction force is $F_{fr}= \mu^i F_n^i f(P)$, where $f(P)=1$ for $P<0.7$ MPa, while for larger pressure $f(P)$ is proportional to the deviation from the linear law of the area fraction. Since this factor is calculated numerically for a finite number of values of $P$, we linearly interpolate the value for a generic pressure in the middle of a range between the calculated ones. The behavior of $f(P)$ obtained is shown in the right panel of figure \ref{fig:AreaFrac}.

Once the friction law is fixed, we tune the friction coefficient so that the dynamic friction coefficient of the nominally flat surface is as close as possible to the experimentally measured value. After preliminary tests, we have fixed the averages of local friction coefficients to $(\mu_s)_m= 0.3$, $(\mu_d)_m = 0.27 $, and their standard deviations to 5\% of the average.

\section{Results}\label{sec4}

\subsection{Measurement of the friction coefficients of flat and patterned surfaces}\label{sec_exp_res}

\begin{figure}[ht]
\hspace{-1.5cm}
\begin{minipage}[c]{.40\textwidth}
\centering
\includegraphics[width=1.6\textwidth]{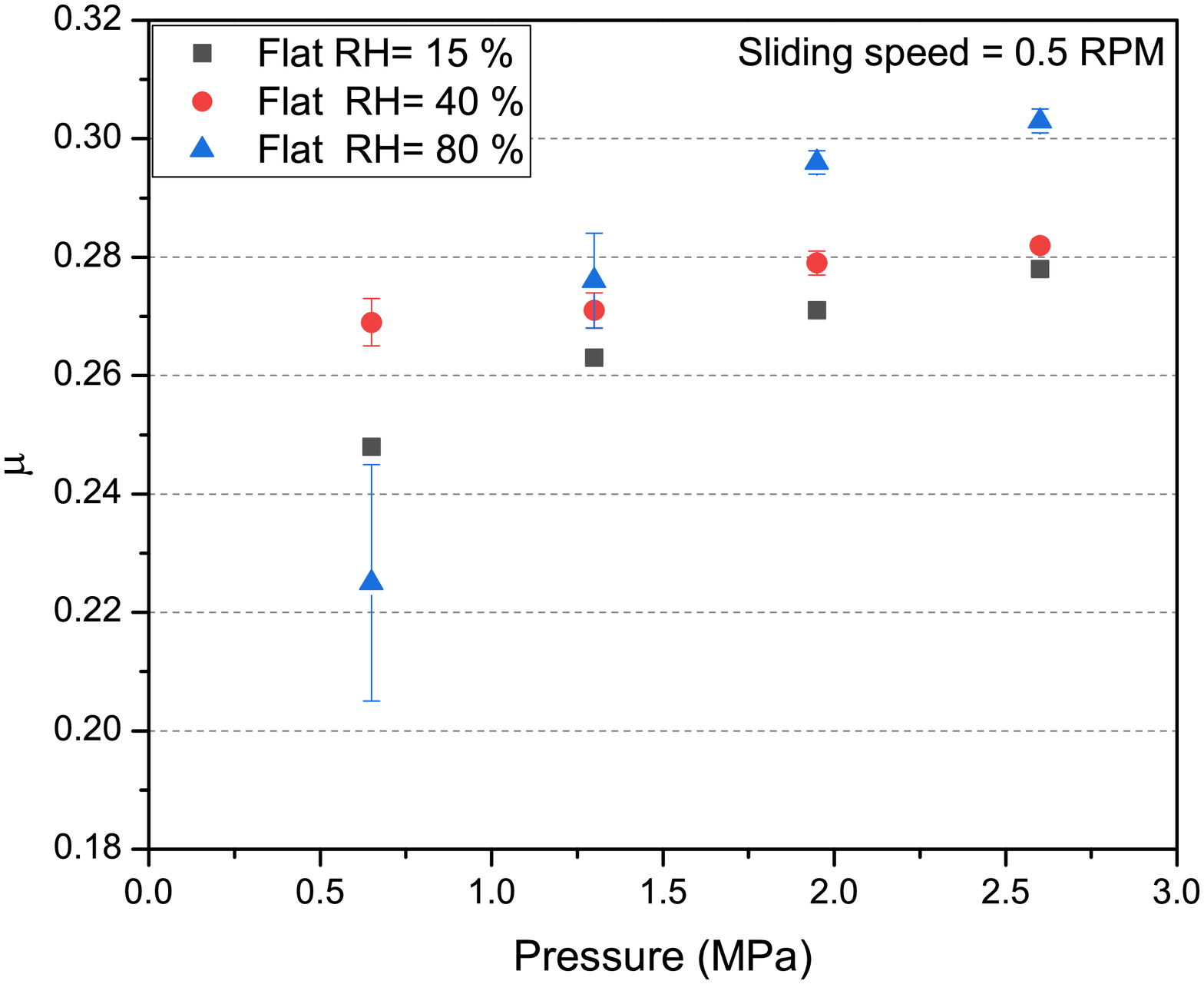}
\end{minipage}%
\hspace{25mm}%
\begin{minipage}[c]{.40\textwidth}
\centering
\includegraphics[width=1.6\textwidth]{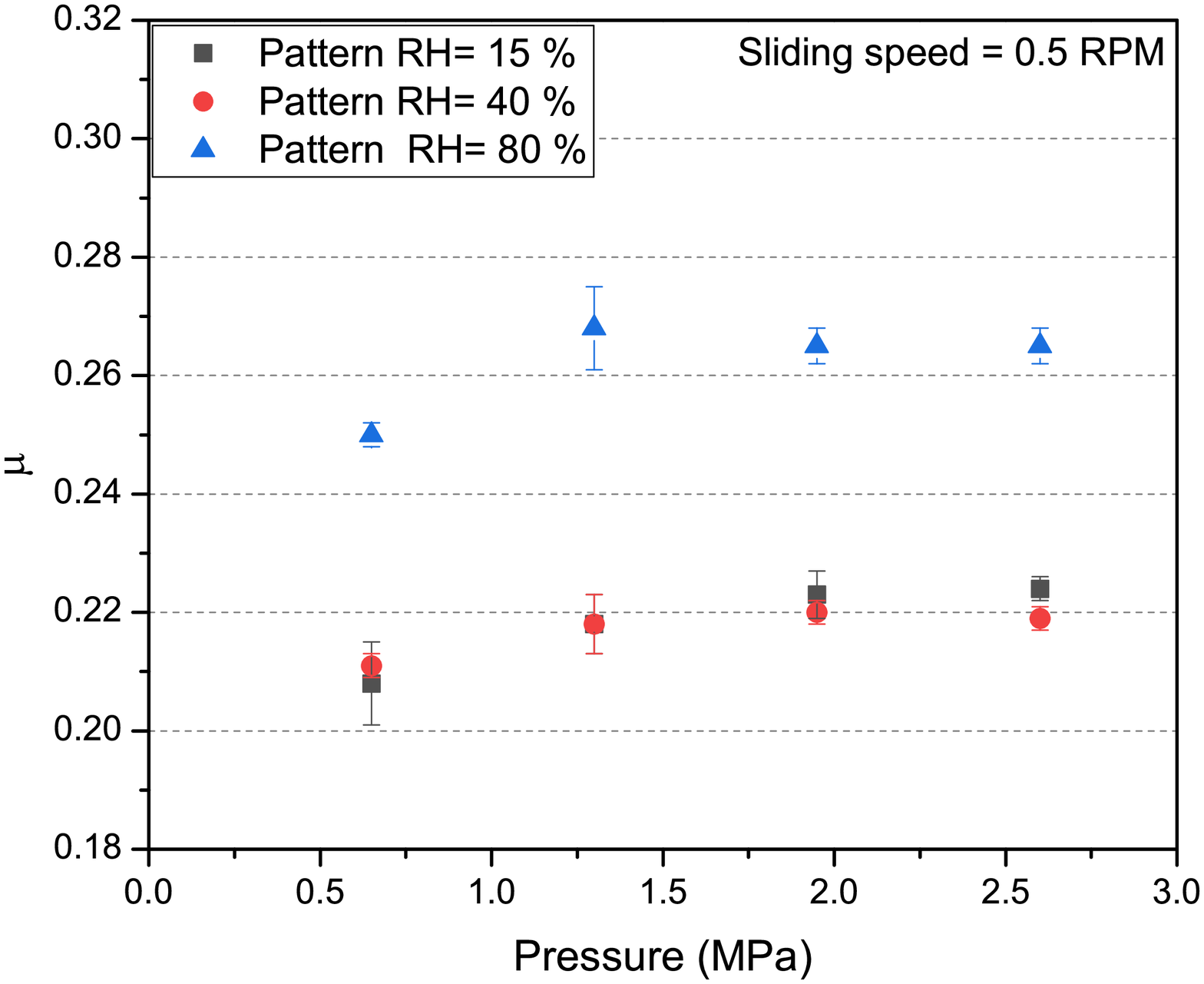}
\end{minipage}
\caption{$\mu$ for flat and patterned samples as a function of normal load and relative humidity (RH).}
\label{fig:FlatPatternCoF}
\end{figure}

The dynamic friction coefficient ($\mu$) is evaluated as a function of sliding velocity, force and relative humidity.
Each test is repeated three times, and the corresponding average and standard deviation is calculated for each measurement (considering $\mu$ values between 10 and 40 seconds). Figure \ref{fig:FlatPatternCoF} shows the output of tests performed on nominally flat and patterned sample as a function of the applied normal force and various humidity, for a  sliding speed of 2 mm/s (0.5 RPM).
Normal force and pressure P are related by the following relationship:

\begin{equation}
P=\frac{F_n}{\pi r^{2}}
\label{eq.NormalForce}
\end{equation}

where r is the radius of the tribometer sample (r = 0.005 m). \\
Results show that the friction coefficient decreases for patterned samples, which is consistent with the literature, e.g. \cite{Maegawa}. In particular, the coefficient of friction of the patterned sample decreases by (18$\pm$3) \% compared to the flat sample at 15 \% relative humidity. This can be explained by stress accumulations at the edges of the cavities, which leads to facilitated detachment. In both cases, the coefficient of friction  increases with the normal force and the relative humidity. All values are reported in table \ref{flatParametri} for the nominally flat case and in table \ref{patternParametri} for the patterned case.
If the pressure increases the coefficient of friction increases, this happens because the real contact area is proportional to the applied pressure. The boundary between the surface is greater if the contact area is wider. Above 2 MPa the coefficient of friction reaches a constant value for the patterned sample due to the saturation of the contact area; for the flat sample this saturation effect  is not reached in the pressure range analyzed. 

\begin{figure}[!h]
\hspace{-1cm}
\begin{minipage}[c]{.40\textwidth}
\centering
\includegraphics[width=1.2\textwidth]{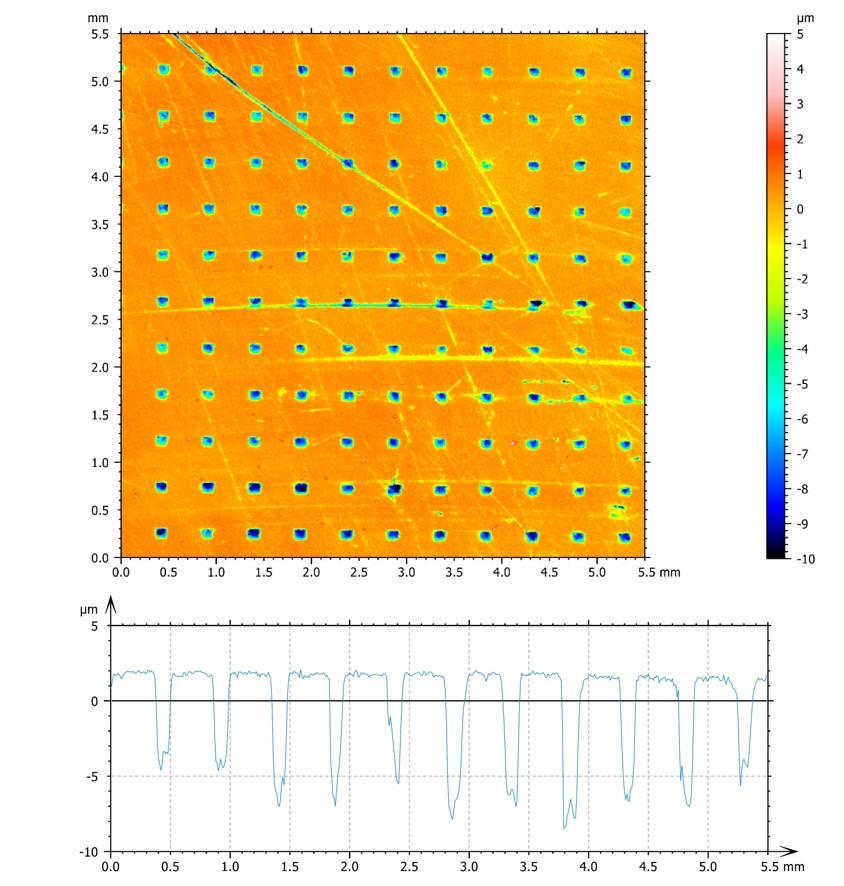}
\end{minipage}%
\hspace{20mm}%
\begin{minipage}[c]{.40\textwidth}
\centering
\includegraphics[width=1.2\textwidth]{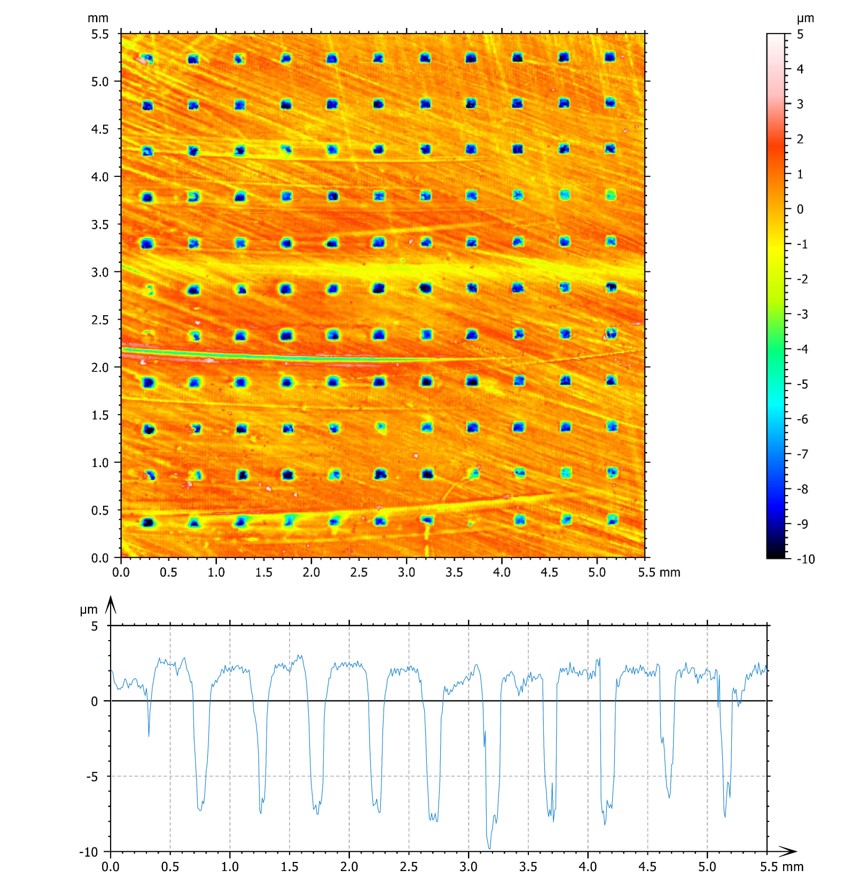}
\end{minipage}
\caption{Profilometry of the patterned sample. (left) before test in the tribometer (Sa=0.789 $\mu$m); (right) after the test (Sa=1.029 $\mu$m). top - altitude map of the patterned surface. bottom - depth profile extracted at height y of 3 mm.}
\label{profilometroPatter}
\end{figure}

Altitude maps show no significant differences before and after the test, both for flat and patterned samples. The cavity profiles are unchanged after the test (Supplementary Material, figure \ref{profilometroPatter}). The surface roughness (Sa) of the samples undergoes slight changes, with the introduction of micro scratches parallel to the direction of sliding.

\begin{table}[h!]
\centering
\begin{tabular}{cccc}
\toprule
{Normal  force - Speed} & {15$\%$  RH}& {40$\%$  RH} & {80$\%$  RH} \\
\midrule
50 N  - 2 mm/s (0.5 RPM) & 0.248$\pm$0.002  &  0.269$\pm$0.004  & 0.225 $\pm$0.02 \\
100 N  - 2 mm/s (0.5 RPM) & 0.263$\pm$0.005  &  0.271$\pm$0.003  &  0.276$\pm$0.008 \\
150 N  - 2 mm/s (0.5 RPM) & 0.271$\pm$0.002   &  0.279$\pm$0.002  &  0.296$\pm$0.002 \\
200 N  - 2 mm/s (0.5 RPM) & 0.278$\pm$0.001  &  0.282$\pm$0.001  &  0.303$\pm$0.002 \\
100 N  - 0.4 mm/s (0.1 RPM) & 0.264$\pm$0.003  &  0.265$\pm$0.007  &  0.30$\pm$0.01 \\
100 N  - 4 mm/s (1 RPM) & 0.287$\pm$0.002  &  0.294$\pm$0.002  &  0.319$\pm$0.005 \\
\bottomrule 
\end{tabular}
\caption{Friction coefficient of a nominally flat sample as a function of normal force, sliding velocity and relative humidity (RH). All measurements are performed at room temperature. }
\label{flatParametri}
\end{table}

\begin{table}[h!]
\centering
\begin{tabular}{cccc}
\toprule
{Normal  force - Speed} & {15$\%$  RH}& {40$\%$  RH} & {80$\%$  RH} \\
\midrule
50 N  - 2 mm/s (0.5 RPM) &  0.208$\pm$0.007 &  0.211$\pm$0.002  &  0.250$\pm$0.002 \\
100 N  - 2 mm/s (0.5 RPM) & 0.218$\pm$0.005  &  0.218$\pm$0.005  & 0.268$\pm$0.007  \\
150 N  - 2 mm/s (0.5 RPM) & 0.223$\pm$0.004  &  0.220$\pm$0.002  &  0.265$\pm$0.003\\
200 N  - 2 mm/s (0.5 RPM) & 0.224$\pm$0.002  &  0.219$\pm$0.001  &  0.265$\pm$0.003\\
100 N  - 0.4 mm/s (0.1 RPM) & 0.200$\pm$0.002 &  0.196$\pm$0.002  &  0.263$\pm$0.003 \\
100 N  - 4 mm/s (1 RPM) & 0.240$\pm$0.002  &  0.238$\pm$0.002  & 0.285$\pm$0.002  \\
\bottomrule 
\end{tabular}
\caption{Friction coefficient of a patterned sample as a function of normal force, sliding velocity and relative  humidity (RH). All the  measurements are performed at room temperature. }
\label{patternParametri}
\end{table}

\clearpage

\subsection{Comparison between experiments and numerical simulations}

\begin{figure}[h!]
\hspace{-1.5cm}
\begin{minipage}[c]{.40\textwidth}
\centering
\includegraphics[width=1.6\textwidth]{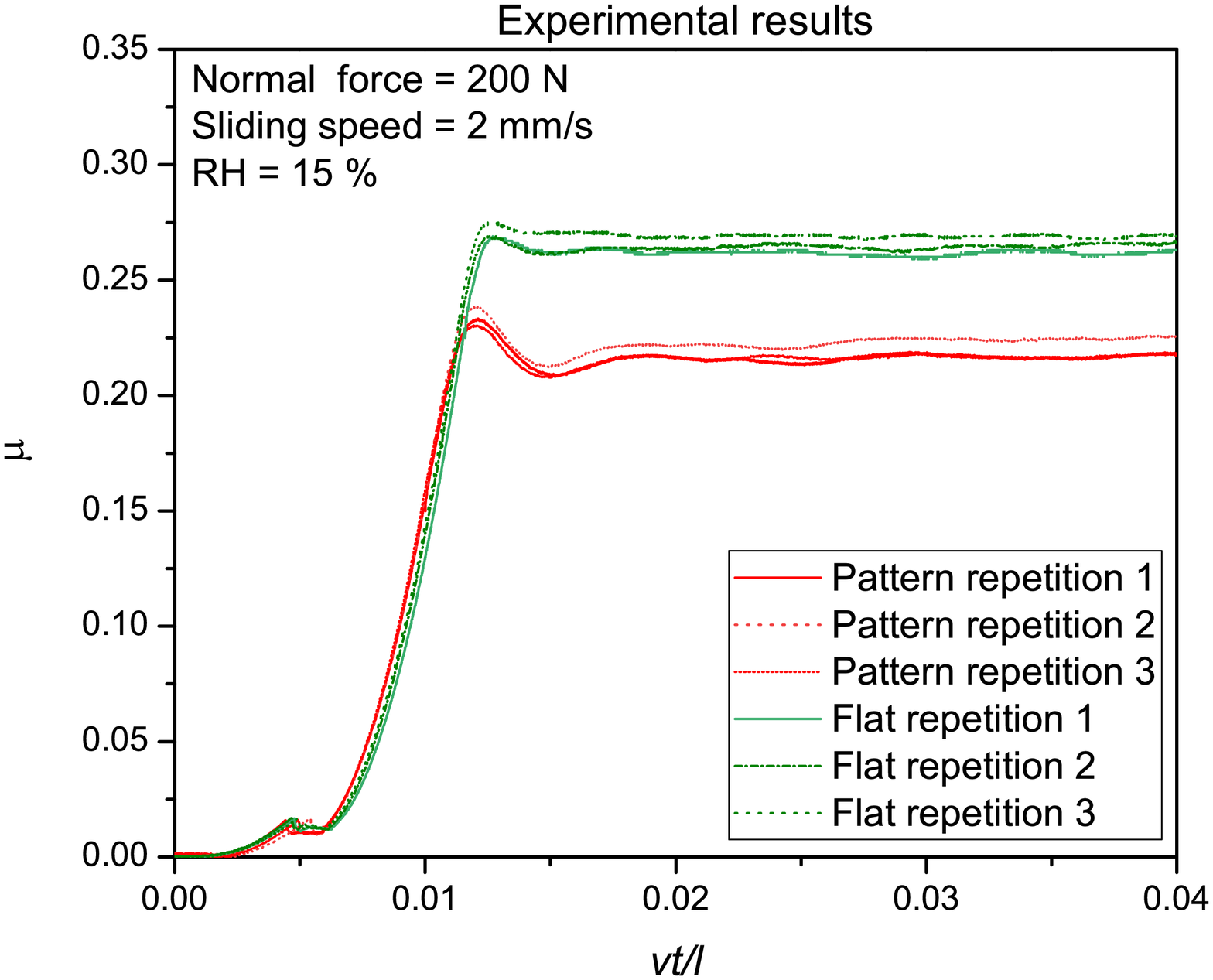}
\end{minipage}%
\hspace{25mm}%
\begin{minipage}[c]{.40\textwidth}
\centering
\includegraphics[width=1.6\textwidth]{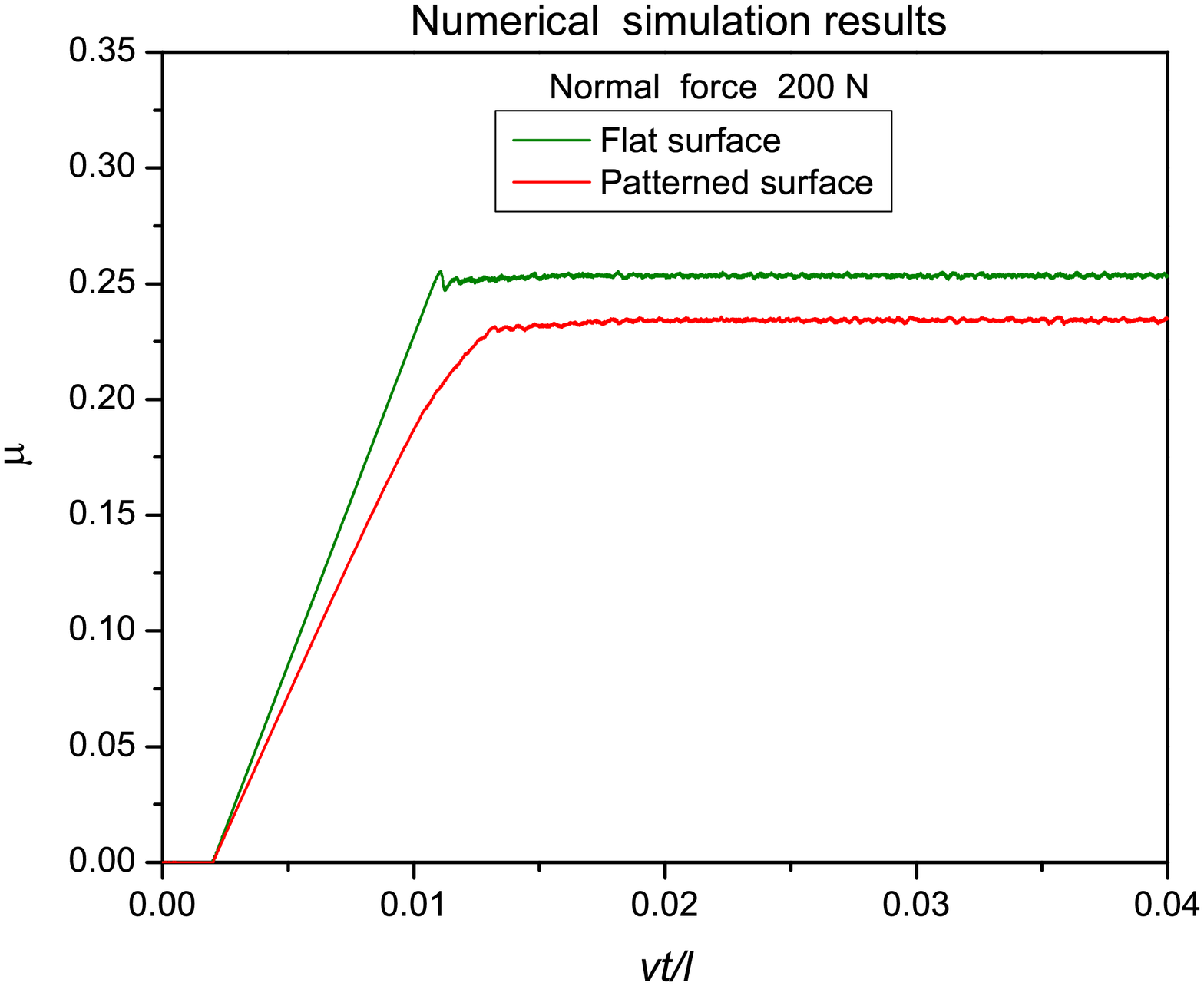}
\end{minipage}
\caption{Comparison of the time evolution of the friction coefficient for the nominally flat and patterned cases, with $v=2$ mm/s, RH=15\%  and total normal force 200 N, obtained from experimental test (left) and numerical simulations (right). Three repetitions are shown for experimental tests, showing good reproducibility.}
\label{fig:FlatPatternTimeEvolution}
\end{figure}

\begin{figure}[h!]
\begin{center}
\includegraphics[scale=0.40]{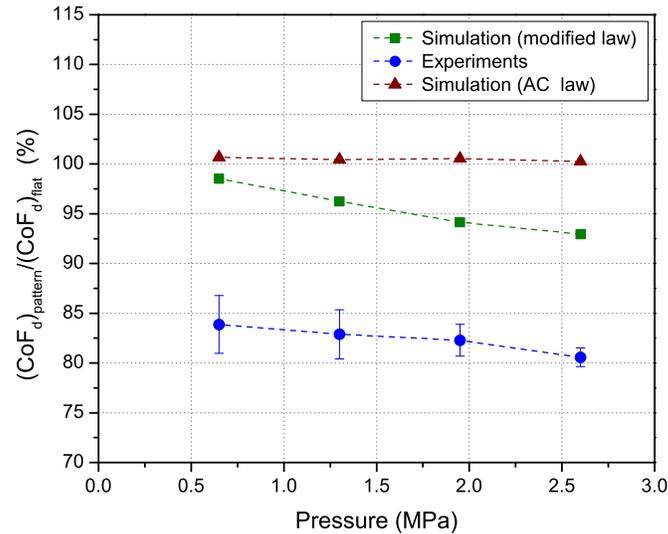}
\caption{ Comparison between simulations (for both an AC law and a modified law) and experimental results for the ratio between patterned and flat dynamic $\mu$ values as a function of pressure.  }
\label{fig:SimResults}
\end{center}
\end{figure}

The numerically simulated curves for the time evolution of the coefficient of friction are qualitatively similar to the experimental behavior \ref{fig:FlatPatternTimeEvolution}. The comparison between the results for the coefficient of friction obtained with the standard AC friction law and the modified law is shown in figure \ref{fig:SimResults}. In the latter case, the macroscopic dynamic friction coefficient, obtained from the time average over the dynamic phase, is smaller in the patterned case compared to the nominally flat case. Due to the patterning, points located near the edges are subjected to larger pressures, but due to the non-linearity of the friction force, they experience a reduced friction force. 

Given the model simplifications, a perfect match of the time evolution curves and the friction coefficients cannot be achieved. However, fitting parameters, e.g. friction coefficients of the local effective friction law, can be tuned to match the macroscopic dynamic friction coefficient for a flat surface determined experimentally. Then, the same parameters can be used for the comparison in the patterned case. The reduction of the macroscopic friction coefficients in the patterned case could never be obtained with a standard AC friction law, regardless of fine parameter tuning. The non-linear behavior of the local friction law, with the non-uniform pressure distribution on the contact area, is a crucial step to improve the results of this model towards experimental observations.

Another point supporting the significance of this mechanism is the decreasing trend of the data set of figure \ref{fig:SimResults} with pressure. This suggests that the effect is enhanced for a larger pressure, as the modified friction law implies. In the presence of mechanisms unrelated to this, e.g. a constant adhesion term \cite{Berardo} or a reduction of the detachment thresholds \cite{Costagliola2}, the ratio between the coefficient of friction for the nominally flat and patterned cases should be constant as a function of the pressure.

The remaining discrepancy between simulations and experiments can be explained by other effects neglected in the model, e.g. stress concentrations in the transversal direction. In this formulation, transversal grooves and square cavities are not distinguishable. Results in \cite{Costagliola2} have shown that differences of a few percent in the coefficient of friction are found between these two cases, and this can explain a part of the discrepancy. 

Plasticity can be another factor. The typical yield strength of low-density polyethylene \cite{Jordan} is larger than the applied pressure of our experiments, but for regions around the cavities, subjected to stress concentrations, plasticity can be a significant effect. A full three-dimensional formulation, taking into account the vertical and planar stress distributions (beyond the scope of this work), can help to quantify the contribution of these different effects.

\clearpage

\subsection{Humidity effect}

\begin{figure}[h!]
\begin{center}
\includegraphics[scale=0.40]{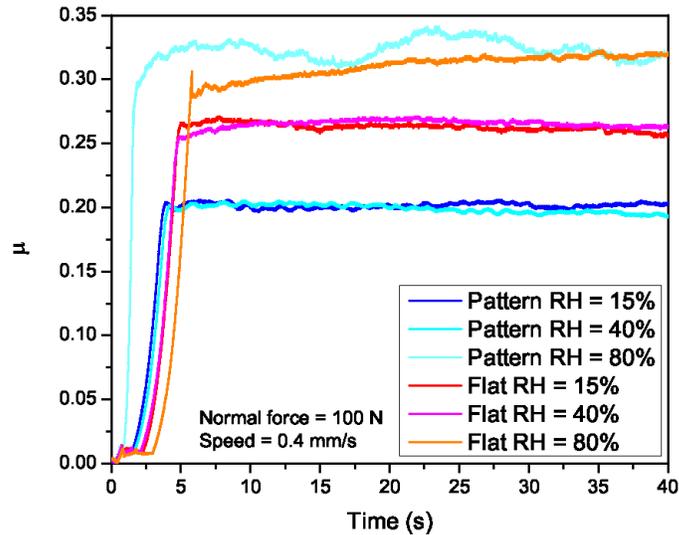}
\caption{Coefficient of friction as function of time for different relative humidity.}
\label{fig:CoFvsTimeHumidity}
\end{center}
\end{figure}

Figure \ref{fig:CoFvsTimeHumidity} shows the coefficient of friction as function of time. The friction curves below 80 \% of relative humidity have negligible differences. The friction curve of the flat sample at 80 \% of relative humidity shows a high value of static friction due to meniscus forces. The pattern sample at high humidity (80 \%) shows friction instability, which could be due to the interaction between the cavities of the patterned surface and the water droplets. 

These results are consistent with those in the literature \cite{Bhushanart}, where the friction coefficient is shown to usually increase with humidity, up to the emergence of the aquaplaning phenomenon. The friction level depends on the surface geometry and the amount of water between the two surfaces. Surface asperities define the creation and the strength of the meniscus bonds when water
is present on the interface. Friction increases with humidity, especially for large relative humidity values (80 \%). It is interesting to note that for a normal load of 50 N, friction of the flat sample decreases at high humidity levels, which means that the surfaces are in an immersed regime \cite{Bhushanart} and so the water film serves as a lubricant. If the normal force is further increased, the excess water is eliminated from the interface. On the other hand, in the patterned sample, the water can be removed more easily from the surface due to the presence of cavities, so that more water is required to obtain an immersed regime compared to the flat sample. The excess water on the surface enters the cavities, so that the effect of relative humidity at 40 \% is negligible in the patterned sample.

\subsection{Sliding velocity dependence}

\begin{figure}
\hspace{-1.5cm}
\begin{minipage}[c]{.40\textwidth}
\centering
\includegraphics[width=1.6\textwidth]{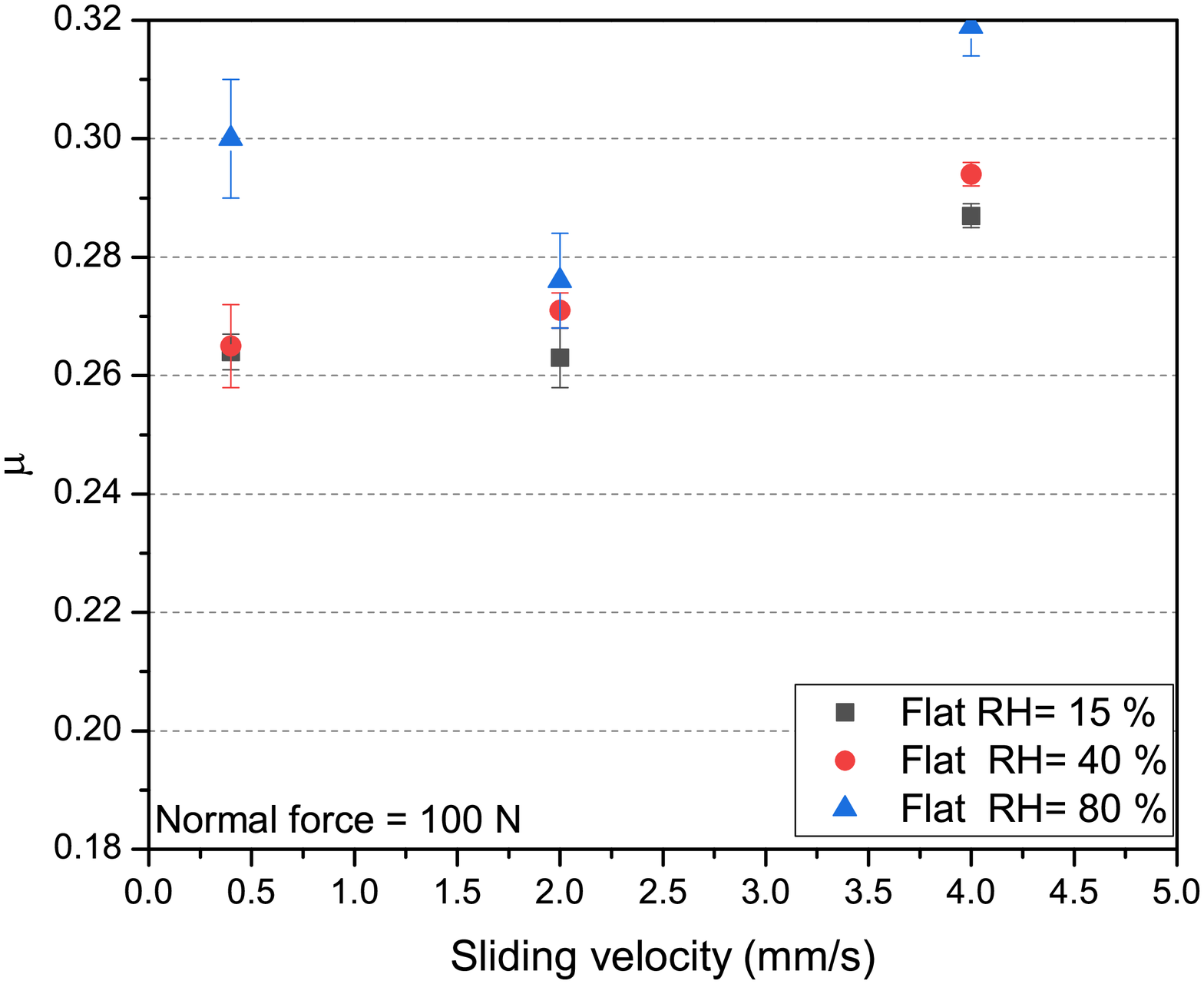}
\end{minipage}%
\hspace{25mm}%
\begin{minipage}[c]{.40\textwidth}
\centering
\includegraphics[width=1.6\textwidth]{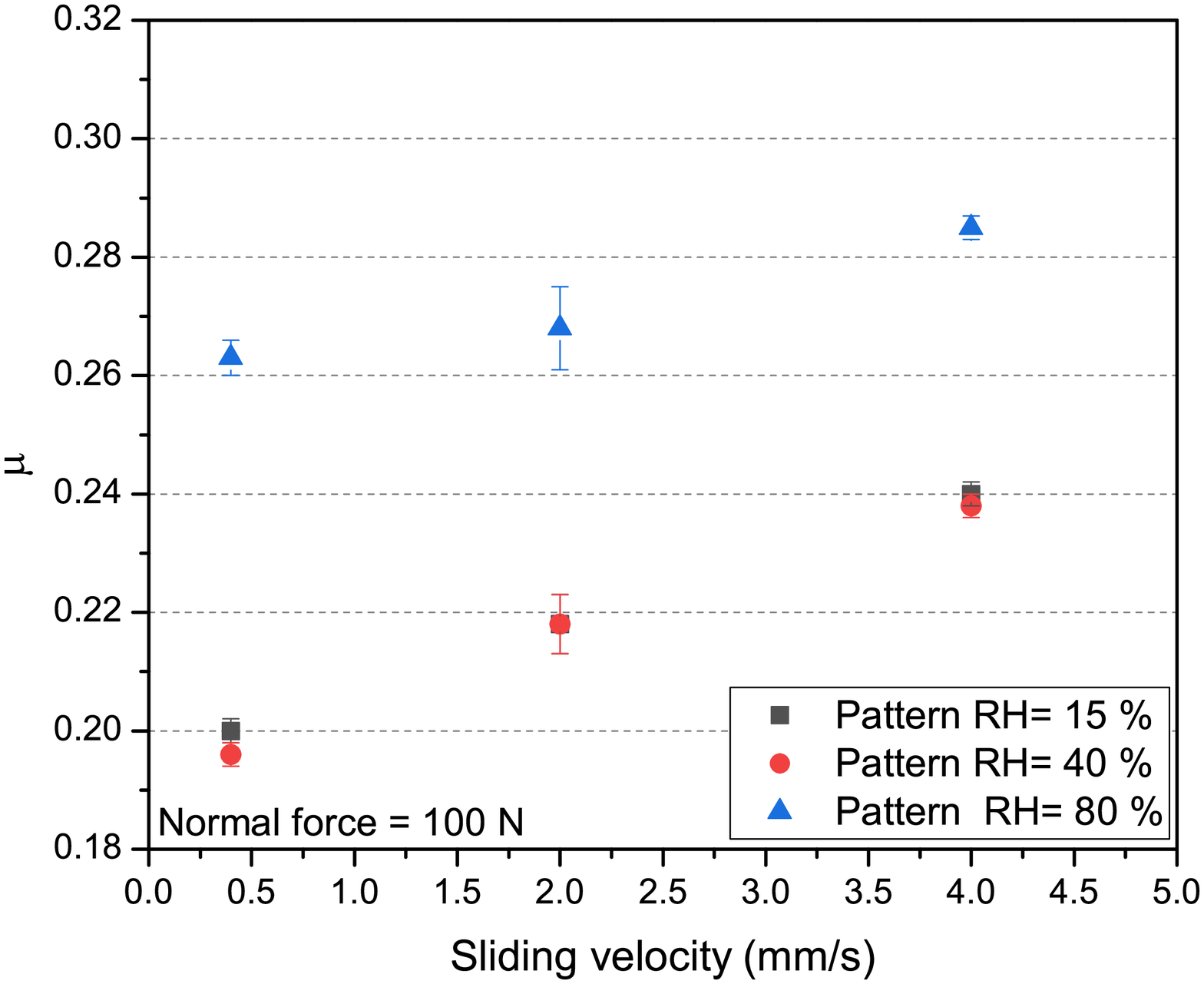}
\end{minipage}
\caption{Experimentally measured $\mu$ for flat and patterned samples as a function of sliding speed. The pressure applied is 1.3 MPa  (normal  force: 100 N). }
\label{fig:CoFSpeed}
\end{figure}

The friction coefficient as a function of the sliding speed displays an increase of 10\% for the flat case and 20\% for the patterned case in the range between 0.4 mm/s and 4 mm/s. This is consistent with experimental results on polymers \cite{Unal}\cite{Bahadur}, which is usually explained with the enhancement of the adhesion between contact asperities caused by the increase of energy dissipation. This can also explain why the relative increase is larger for the patterned sample, since in this case the local pressure on contact points is larger. The only exception to this trend is for the flat sample at 80 \% humidity, but here statistical uncertainties are also larger. There are no significant differences between the relative humidity of 15 and 40 \%.  All tests were carried out for a normal load equal to 100 N (corresponding to 1.3 MPa of pressure). Figure \ref{fig:CoFSpeed} shows friction coefficient values as a function of sliding speed.  

\begin{figure}[h!]
\begin{center}
\includegraphics[scale=0.40]{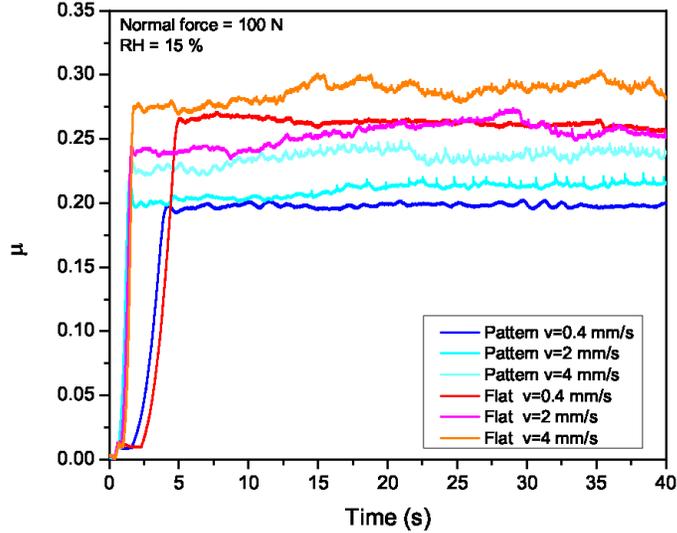}
\caption{Comparison of the time evolution of the friction coefficient for the nominally flat and patterned cases, with $F=100$ N, RH=15\%.}
\label{fig:Temporalevolution}
\end{center}
\end{figure}

Figure \ref{fig:Temporalevolution} shows the temporal evolution of coefficient of friction. The static phase for the speed of 0.4 mm/s lasts longer. The dynamic coefficient of friction for speeds larger than 0.4 mm/s displays larger fluctuations during the test, which can be ascribed to larger instabilities and stick-slip events.

\subsection{Stick-Slip}

\begin{figure}
\hspace{-1.5cm}
\begin{minipage}[c]{.40\textwidth}
\centering
\includegraphics[width=1.4\textwidth]{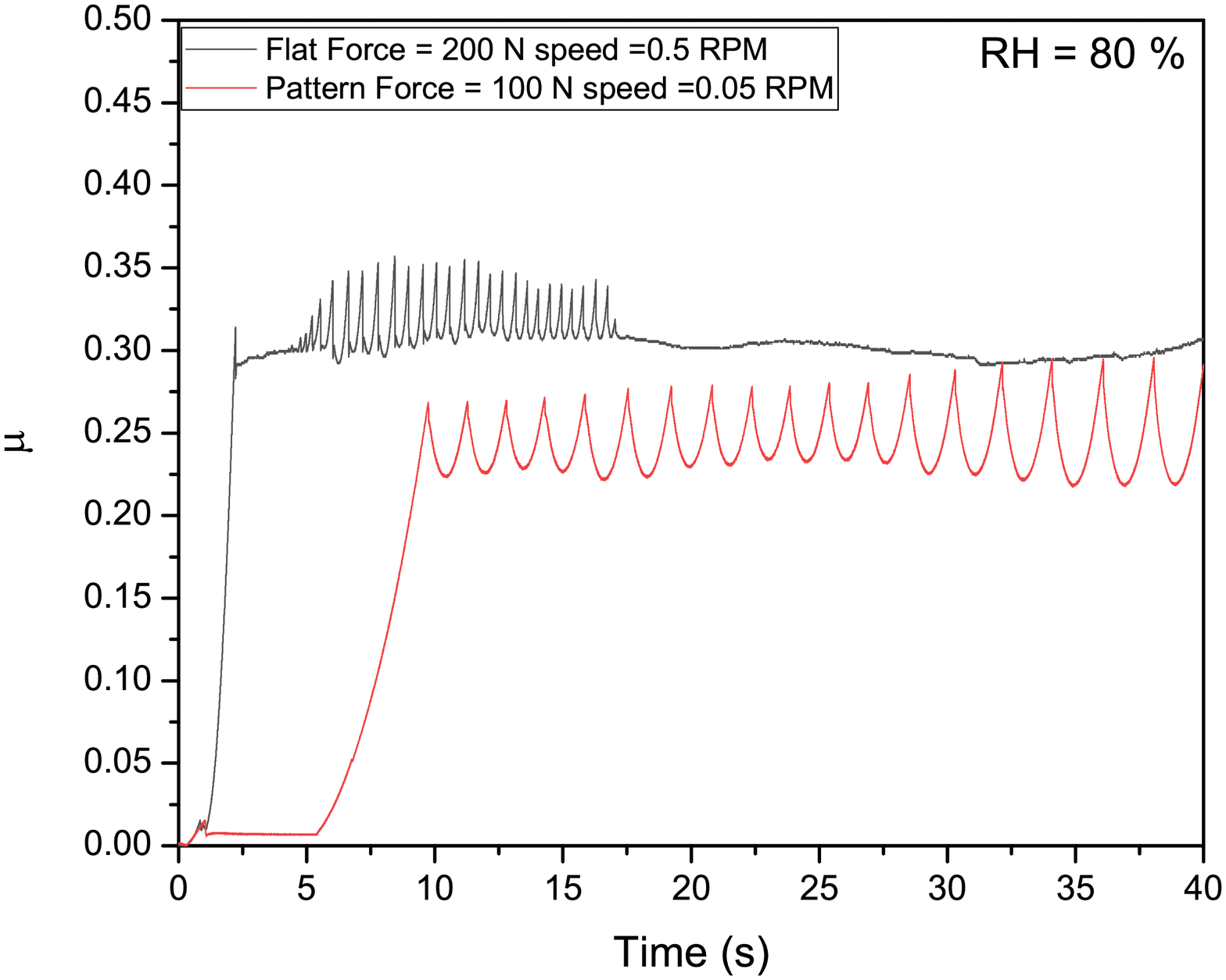}
\end{minipage}%
\hspace{30mm}%
\begin{minipage}[c]{.40\textwidth}
\centering
\includegraphics[width=1.3\textwidth]{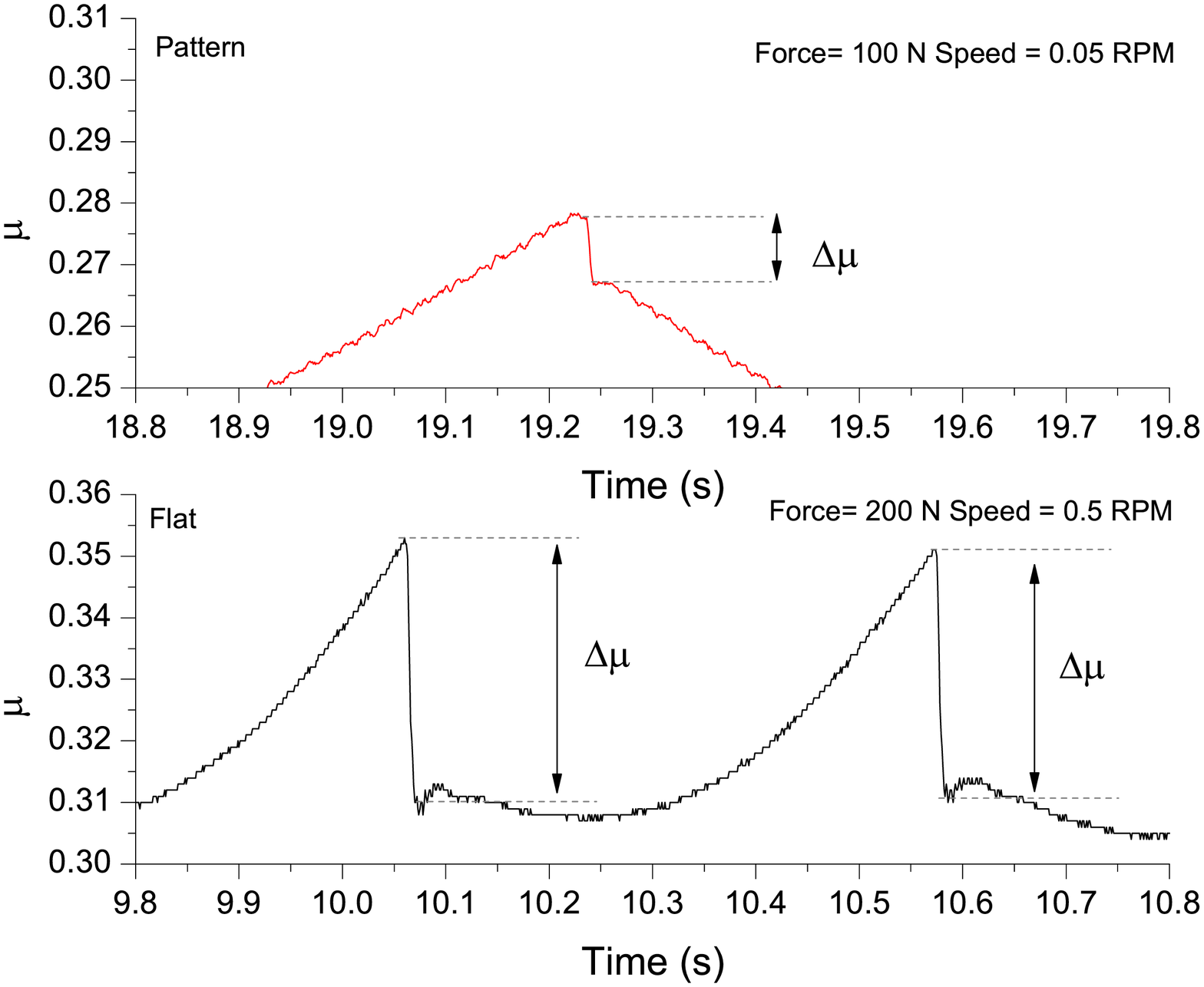}
\end{minipage}
\caption{(left) Stick-slip phenomena in tribometer tests. (right) Close-up of stick-slip events. The stick-slip amplitude decreases for the patterned sample.}
\label{fig:Stick-Slip}
\end{figure}

The onset of a stick-slip regime occurs in the high humidity regime (80 \% of relative humidity) due to the meniscus forces that are created between the surfaces. Stick-slip occurs at different normal loads and sliding velocity values for flat and patterned samples. An important effect of sample patterning is that in addition to reducing the friction coefficient, it also influences the stick-slip phenomenon. This occurs at different loads and sliding speeds for flat and patterned samples. The patterned sample displays stick-slip at lower loads and sliding speeds. This is due to the interplay between stress concentrations on the surface and the influence of humidity. The amplitude of stick-slip ($\Delta \mu$) is smaller for the patterned sample, indicating a smaller difference between static and dynamic coefficient of friction. Figure \ref{fig:Stick-Slip}  (right) shows the difference between $\Delta \mu$ of  pattern and flat  sample.
Surface geometry modifies the distribution of the water on the surface, with the consequent modification of adhesion forces. Using patterned surfaces can provide a means to modify the parameter ranges in which stick-slip takes place without changing the chemical composition of the surfaces, which can be useful in many applications, including groan and squeal noise reduction. Reducing the amplitude of stick-slip can also be useful to reduce the mechanical vibrations generated during friction.

\clearpage

\clearpage
\section{Conclusions}\label{sec5}

In this article, we examine how the frictional properties of polymeric samples can be modified with surface patterning. We find that the dynamic friction coefficient decreases with respect to that of a non-patterned surface and that the stick-slip behavior also changes in the presence of surface cavities. Experimental results for dynamic friction are compared with the predictions obtained by a simplified two-dimensional spring-block model in the vertical plane. We find a qualitative agreement, provided that the AC local friction law is modified to take into account the non-linear behavior of the real contact area for larger applied pressures. This correction is obtained by calculating the real contact area by using a Boundary Element Method solver with experimental surface profiles as input. With this modification, the model correctly reproduces the experimental trends, while underestimating the difference between nominally flat and patterned samples. This discrepancy can be attributed to the two-dimensional approximation adopted in simulations or to neglected plasticity effects, which will be addressed in future works with a more realistic three-dimensional approach. \\
Experimental results also show that friction increases with humidity, and that the onset of stick-slip events occurs in the high humidity regime. \\
Surface patterning is a potentially attractive method to modify in a controlled manner the friction coefficient of a material. The novelty of the approach presented herein is to provide, through a coupled numerical-experimental approach, a method to predict the effective frictional behaviour of an arbitrarily patterned surface. For a given material, experiments allow the validation of a spring-block model and determination of local (microscale) friction coefficients for a given material. The use of experimentally measured surface profiles can then be used as input for the model, based on a nonlinear AC law, providing increased predictive power for nontrivial surface structures. These can be used to achieve controlled frictional properties in the required ranges of normal force, sliding speed and humidity for practical applications.

\clearpage

\section{Supplementary material}\label{sec_suppl}

\begin{figure}[!h]
\hspace{-1cm}
\begin{minipage}[c]{.40\textwidth}
\centering
\includegraphics[width=1.2\textwidth]{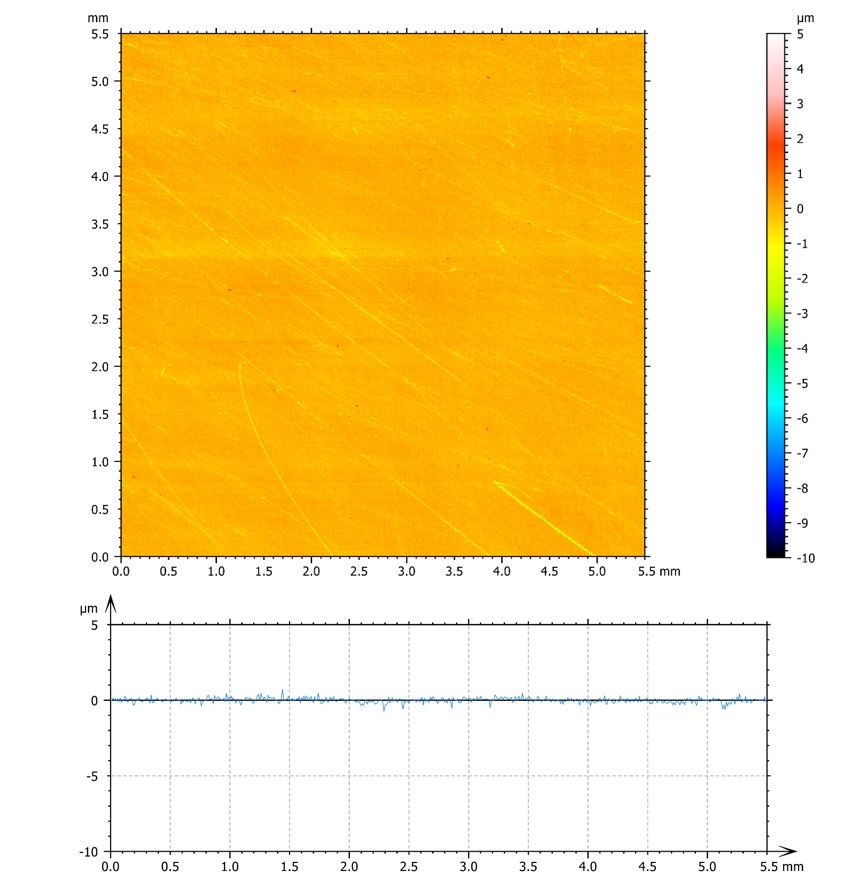}
\end{minipage}%
\hspace{20mm}%
\begin{minipage}[c]{.40\textwidth}
\centering
\includegraphics[width=1.2\textwidth]{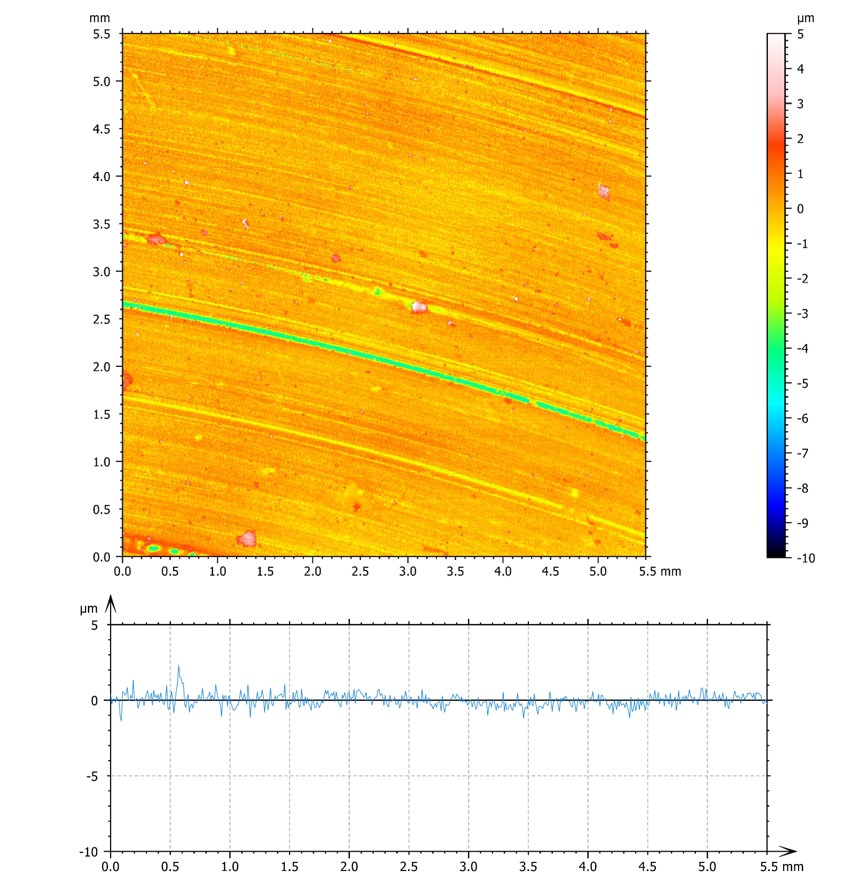}
\end{minipage}
\caption{Topology of a flat sample.  (left) altitude map before test in the tribometer (Sa=0.158 $\mu$m), (right) after the test. top - altitude map of the patterned surface (Sa=0.426 $\mu$m). bottom - depth profile extracted at height y of 3 mm.}
\label{profilometroFlat}
\end{figure}

\section{Acknowledgements}

FB and NMP are supported by H2020 FET Open “Boheme” grant No. 863179. GC has received funding from the European union's Horizon 2020 research and innovation programme under the Marie Sklodowska-Curie grant agreement no. 754462.
All authors would like to thank ITT friction technologies and Bruker for the development tests of the UMT TriboLab tribometer. \\
AS, EV, SB, AP carried out this research  within the framework of the ITT-UniTo JointLab initiative.

\newpage
\null

\end{document}